\documentclass{aa}  

\usepackage{graphicx}
\usepackage{txfonts}
\begin{document}

   \title{Taking advantage of photometric galaxy catalogues to determine the halo occupation distribution}

   %\subtitle{I. Overviewing the $\kappa$-mechanism}

   \author{F. Rodriguez\thanks{E-mail:facundo@oac.unc.edu.ar (FR);}
          \inst{1}
          \and
          M. Merchán\inst{1}
          \and
          M. A. Sgr\'o\inst{1}
          }

   \institute{Instituto de Astronom\'ia Te\'orica y Experimental (UNC-CONICET), 
              Observatorio Astron\'omico de C\'ordoba, Laprida 854, C\'ordoba X500BGR, Argentina\\
              %\email{wuchterl@amok.ast.univie.ac.at}
              }

   \date{Received February 3, 2015}

% \abstract{}{}{}{}{} 
% 5 {} token are mandatory
 
  \abstract
   {Halo occupation distribution (HOD) is a powerful statistic that allows the  study of  several aspects of the matter distribution in the universe, 
    such as evaluating semi-analytic models of galaxy formation or imposing constraints on cosmological models.
   Consequently, it is important to have a reliable method for estimating this statistic, taking full advantage of the available information on current and future galaxy surveys.
   }
   {The main goal of this project is to combine photometric and spectroscopic information using a discount method of background galaxies
   in order to extend the range of absolute magnitudes and to increase the upper limit of masses in which the HOD is estimated. 
   We also evaluate the proposed method and apply it to estimating the HOD on the Sloan Digital Sky
   Survey Data Release 7 (SDSS DR7) galaxy survey.}
   {We propose   the background subtraction technique to mel information provided by spectroscopic galaxy groups and photometric survey of galaxies. 
   To evaluate the feasibility of the method, we implement the proposed technique on a mock catalogue built from a semi-analytic model of galaxy formation. 
   Furthermore, we apply the method to the SDSS DR7 using a galaxy group catalogue taken from spectroscopic version and the corresponding photometric galaxy survey.}
   { We demonstrated the validity of the method using the mock catalogue.
   We applied this technique to obtain the SDSS DR7 HOD in absolute magnitudes ranging from $M=-21.5$ to $M=-16.0$ and masses up to $\simeq 10^{15} M_{\odot}$ throughout this range. 
   On the brighter extreme, we found that our results are in excellent agreement with those obtained in previous works.}
   {}


   \maketitle
%
%________________________________________________________________

\section{Introduction}

   The current paradigm assumes that galaxies form by baryon condensation within the potential wells
   defined by the collisionless collapse of dark matter haloes \citep{White&rees78}, but the 
   diversity of astrophysical mechanisms involved in galaxy formation do not allow us to
   determine how galaxies occupy haloes.
   Clustering studies are  powerful tools used to connect galaxies to their unseen dark matter haloes and indicate
   how these were assembled.

   Recently, with large galaxy surveys, the halo occupation distribution (HOD) and conditional luminosity
   function (CLF) are intensively investigated as methods for studying galaxy clustering.
   The two methods are closely related, but CLF considers the average number of galaxies with luminosity 
   L $\pm$ dL/2 that reside in a halo of mass $M_h$ (e.g. \cite{vandenBosch03-2}; \cite{Yang03}; \cite{Vale04}; \cite{vandenBosch07}; \cite{Vale06})
   and HOD describes the probability distribution $P(N|M_h)$ that a virialized halo of mass $M_h$ contains $N$ galaxies 
   with some specified characteristic (e.g. \cite{Jing98}; \cite{Jing98-2}; \cite{Ma00}; \cite{Peacock00}; 
   \cite{Seljak00}; \cite{Scocciamarro01}; \cite{Berlind02}; \cite{Cooray02}; \cite{Berlind03}; 
   \cite{Zheng05};\cite{Yang08}). 

   Many authors pointed out the importance of HOD when 
   describing the relation between galaxies and their hosting dark matter haloes 
   (e.g. \cite{Yang07}; \cite{Yang08}). This description is useful in order to constrain
   models of galaxy formation and evolution (e.g. \cite{Berlind03}; \cite{Kravtsov04}); it
   can provide constraints to cosmological models (\cite{vandenBosch03}; 
   \cite{Zheng07}); and it allows  the contribution of central and 
   satellite galaxies to be studied separately 
   (\cite{Kravtsov04}; \cite{Cooray05}; \cite{Zheng05}; \cite{White07}; \cite{Yang08}). 
   The last property contains important information about hierarchical clustering because 
   the number of satellite galaxies around the central galaxy in a dark matter halo 
   provides insights into the formation of massive haloes from smaller ones and can be used to constrain 
   cosmological parameters. Furthermore, as pointed out by \cite{Wang11}, the overprediction of satellite galaxies by numerical simulations called
   ``missing satellite galaxies'' is a particularly striking example of how satellite galaxy studies can constrain cosmology
   and galaxy evolution. 
   This gives valuable information that can be compared with simulations and  used to constrain semi-analytic models.

   As indicated by \cite{Liu2011}, a common method for overcoming the lack of 
   spectroscopic information for satellite galaxies is 
   to count candidate satellites around a bright host in photometric data and to 
   statistically correct for the contribution of foreground and background 
   objects. Consequently a lower limit of apparent magnitudes could be accomplished by employing the enormous volume of
   photometric data together with a background subtraction approach, which is an 
   efficient method for studying the galaxy population on a statistical basis 
   (\cite{Lares2011}). Some examples of satellite galaxies at fainter magnitudes combining a redshift survey with a photometric survey
   are \cite{Holmberg69}, \cite{Phillipps87}, \cite{Lorrimer94},
   \cite{Smith04}, \cite{Chen06}, \cite{Lares2011}, \cite{Guo11}, and \cite{Tal12}).

   Most studies have taken advantage of the large volume of data provided by the Sloan Digital Sky Survey (SDSS; \cite{York00}) and
   use both spectroscopic and photometric data to study the luminosity function and number density of satellite galaxies 
   (e.g. \cite{Chen06}; \cite{Lares2011}; \cite{Guo11}; \cite{Tal12}) but there are no similar works about HOD. 

   In this work we focus on developing the background subtraction method as a technique 
   to determine the HOD. Given the depth of the Sloan photometric catalogue,
   this technique will allow us to expand the range of limiting absolute magnitude;
   at the same time, more distant galaxy groups will be included. 
   This feature will result in group samples with a wider mass range.

   The background subtraction method applied to the HOD determination 
   has not been widely used and is frequently criticized for its inability to
   distinguish between the local and overlying overdensities.
   Consequently, to demonstrate the correct performance of our implementation,
   we have constructed a synthetic catalogue from the semi-analytic model of galaxy formation
   developed by \cite{Guo10} applied to the Millennium I Simulation.

   Using galaxy groups provided by \cite{Yang12} taken from  the spectroscopic version of the SDSS DR7 survey
   and galaxies from the corresponding photometric galaxy catalogue,
   we estimate the HOD for different limiting absolute magnitudes
   and the associated parameterizations according to \cite{Zheng05}.
   These results are compared with the results of \cite{Yang08}, who -- based on the same formalism and using the SDSS data release 4 group catalogue --
   determine the parameters of the HOD for different absolute magnitudes using volume limited samples.

   This paper is organized as follows. In section \ref{BS} we explain the procedure.
   Section \ref{data}  describes real and simulated data. The test of the method is 
   presented in section \ref{test}. Section \ref{I} shows the HOD of
   the SDSS DR7 estimated using our method. Finally, our results are summarized in 
   section \ref{C}.
%__________________________________________________________________

\section{Background subtraction method to estimate the HOD}
\label{BS}
   For the background subtraction procedure it is assumed that large scale galaxy distribution 
   is uniform, while groups are local overdensities. As is explained below, our 
   method statistically estimates the number of galaxies that are in front of 
   and/or behind the group (hereafter background galaxies), but, in projection, 
   seem to belong to it.
   Given that this technique allows  galaxy systems taken from 
   spectroscopic surveys to be combined with catalogues without redshift information,
   it is possible to estimate the HOD in a wider range of magnitudes and at the same time 
   improve the statistics.
   The main contribution of this method is the ability to determine the HOD
   of satellite galaxies in fainter magnitudes than those studied until now.
   
   The background subtraction technique involves counting objects in a region 
   where there is a known signal superposed on an uncorrelated 
   noise, and subtracting a statistical estimation of that noise. 
   In our case, the signal is conformed by the overdensities where the galaxy groups 
   lie, meanwhile the noise is associated with the background galaxies that do 
   not belong to the group.
   The proposed method require two catalogues that share the same sky area, one for groups 
   and another for galaxies.
   For each group the method requires mass, redshift, angular positions, and a characteristic radius, 
   whereas for galaxies only angular positions and apparent magnitudes are necessary.

   Given that only photometric information is available, to estimate the HOD
   we compute the absolute magnitude assuming that all galaxies are located
   at the same redshift of the group $z_{gr}$,
   \begin{equation}
    M=m-25-5\log (d_L(z_{gr}))
   ,\end{equation}
   where $d_L(z_{gr})$ is the luminosity distance in Mpc at $z_{gr}$.
   
   Then we define $n_i$ as the number of galaxies that have an absolute
   magnitude $M\leq M_{lim}$ within a circle centred in each group with a radius 
   determined by the corresponding projected characteristic radius on the sky. 
   The value of $M_{lim}$ sets the upper limit for the range of absolute magnitudes 
   where the HOD will be estimated.   

   Because it is not possible to distinguish between
   galaxies belonging to the group and background galaxies,
   we need some way to know the local number of background galaxies in order to 
   discount it from $n_i$. 
   The contribution of the background galaxies  to $n_i$ cannot be determined straightforwardly; 
   consequently, a statistical method is required.
   Taking into account the hierarchical behaviour of the large scale structure, 
   it is known that a given overdensity is always immersed in a bigger structure;   
   therefore, we estimate the background contribution by counting the number of galaxies, 
   $n_o$, that meet our selection criteria  ($M\leq M_{lim}$) in an annulus around 
   each galaxy group instead of taking the average density over the whole catalogue.
   The number of galaxies $N$ can be estimated subtracting the local background 
   density multiplied by the projected area for each group (i.e. the background 
   contribution to this area),
   \begin{equation}
   N = n_i - \frac{n_o}{A_o} A_i\,,
   \end{equation}
   where $A_i$ and $A_o$ are the projected areas of a circle centred on the group and
   the corresponding annulus, respectively.
   
   Since the method has no restrictions on the choice of groups, we should be careful 
   when choosing the samples used in each measurement.
   Specifically, low mass groups cannot be detected at high redshifts because they typically contain 
   weaker galaxies, while high mass groups can be observed in the whole range of redshift.   
   To consider this fact, we perform different cuts in redshift for each mass range such that at 
   least the brightest galaxy in each group in this mass range is visible.
   
   Finally, groups are binned in mass intervals to obtain the HOD as the average of 
   $N$, $\langle N|M_h\rangle $.
   It should be taken into account that the groups used in this method should be near enough 
   so that a galaxy with absolute magnitude equal to $M_{lim}$ at the group redshift has 
   an apparent magnitude brighter than the limiting apparent magnitude of the photometric catalogue. 
   
%__________________________________________________________________
  
   \section{Data}
   \label{data}
   To test the described method we  take advantage of simulated data, while 
   to obtain an estimation of the HOD we  use the SDSS DR7 galaxy catalogue.
   In this work, as in Millennium I Simulation, we adopt the cosmological parameters
   taken from WMAP1 data \citep{Spergel03}, i.e.  a flat cosmological model with a 
   non-vanishing cosmological constant ($\Lambda$CDM): $\Omega _m$=0.25, $\Omega _b$=0.045, 
   $\Omega _{\Lambda }$=0.75, $\sigma _8$=0.9, $n$=1, and $h$=0.73.

   \subsection{SDSS Catalogues}
   \label{2}

   The Sloan Digital Sky Survey  
   is a photometric and spectroscopic survey constructed 
   with a dedicated $2.5m$ telescope at Apache Point Observatory in New Mexico. 
  The main advantage of this work is to combine photometric and spectroscopic data,
   and so we will take advantage of the full set of data.
   In particular, we employ the Legacy footprint area which covers more than 8400 $deg^2$ in five optical bandpasses. 
   Imaging data have 230 million objects with apparent magnitudes in the range of  the $r$-band of less than $22.2$,
   and spectroscopic data provide approximately 900,000 
   spectra of galaxies with redshift measurements up to $z\lesssim 0.3$ and 
   an upper apparent magnitude limit of $17.77$ in the $r$-band \citep{Abazajian09}.

   The groups used in our analysis are taken from the SDSS  galaxy group
   catalogue of \cite{Yang07}, constructed  using the adaptive
   halo-based group finder  presented in \cite{Yang05},  
   but updated  to DR7 \citep{Yang12}. Unlike the traditional friend-of-friends (FOF) methods 
   (e.g. \cite{Merchan05}), a property 
   of this group finder is the ability to identify systems even when they have only a 
   single member, which increases the number of groups and at the same time  allows 
   a wider dynamic range of masses. We also use this group catalogue because it will 
   allow us to compare our results with those obtained previously by the same author.

   \subsection{Simulated data}

   In order to test our method, we construct a mock catalogue using synthetic galaxies 
   extracted from a semi-analytic model of galaxy formation applied on top of the Millennium 
   Run Simulation I.

   The Millennium Simulation is a cosmological N-body simulation \citep{Springel05},
   which offers high spatial and time resolution within a large cosmological volume.
   This simulation evolves more than 10 billion dark matter particles in a $500 h^{-1}Mpc$ 
   periodic box, using a comoving softening length of $5 h^{-1}kpc$. 
   Given that the Millennium Simulation is dark matter only, it can be populated 
   with galaxies using a semi-analytic galaxy formation model. 
   For our synthetic catalogue we use the model developed by  \cite{Guo10} publicly 
   available at the Millennium Database 
   \footnote{http://gavo.mpa-garching.mpg.de/MyMillennium}. 
   We also take from this database the 
   information about the dark matter halo where galaxies reside.

   To build our mock catalogue, we place the observer at the origin of the
   Millennium simulation box and, taking account of the periodicity, we 
   simply repeat the simulated volume as many times as necessary to reproduce the 
   volume of the photometric version of the SDSS DR7 catalogue.
   The redshifts are calculated from the distances to the observer taking into account
   the distortion produced by proper motion and the cosmological parameters used in the simulation.
   The flux limited selection of the mock catalogue is then obtained by imposing the 
   same upper apparent magnitude threshold that affects the SDSS catalogue.
   In order to determine the angular selection function of the survey, we use the mask 
   described in section \ref{I}. In addition, a mock group sample is produced from the mock catalogue 
   keeping the original dark matter halo membership used in the simulation.
   
%__________________________________________________________________

   \section{Test of the method}
   \label{test}
   The aim of this section is to demonstrate the effectiveness of the proposed technique.
   For this purpose, we use the mock catalogue described above where we know in detail the
   population of galaxies in dark matter haloes, which can be directly compared to the results obtained with our procedure.
   
   As  is known, flux limited samples do not provide enough information to determine a representative 
   size of groups because the number of visible members depends on the distance.
   Consequently, we use the semi-analytic  model to set a relation between the characteristic 
   radius of groups and their masses, which can be estimated using the visible members.
   For our purposes, we define the characteristic radius as the
   distance from the centre of the host halo to the most distant galaxy member. 
   Assuming a linear behaviour between mass and volume, we propose the  
   relation
   \begin{equation}
   r_c = C_1M_h^{\frac{1}{3}}-C_2
   ,\end{equation}
   where $r_c$ is the characteristic radius and $M_h$ is the host halo mass. 
   The parameters $C_1$ and $C_2$ are determined by fitting the semi-analytic data 
   using a simple chi-square minimization technique, obtaining $C_1=4.31\times 10^{-5} Mpc$ $M_{\odot}^{-\frac{1}{3}}$ 
   and $C_2=-0.33 Mpc$.
   To compute the background contribution as described in section \ref{BS}, the internal and external radius of the annulus
   are chosen as $r_c+1Mpc$ and $r_c+2Mpc$, respectively. 
   We found that the results are not sensitive to the size of this annulus, allowing us to choose a 
   compromise  between the influence of the local overdensity and statistical robustness.
   
\begin{figure}
   \centering
   \includegraphics[width=\hsize]{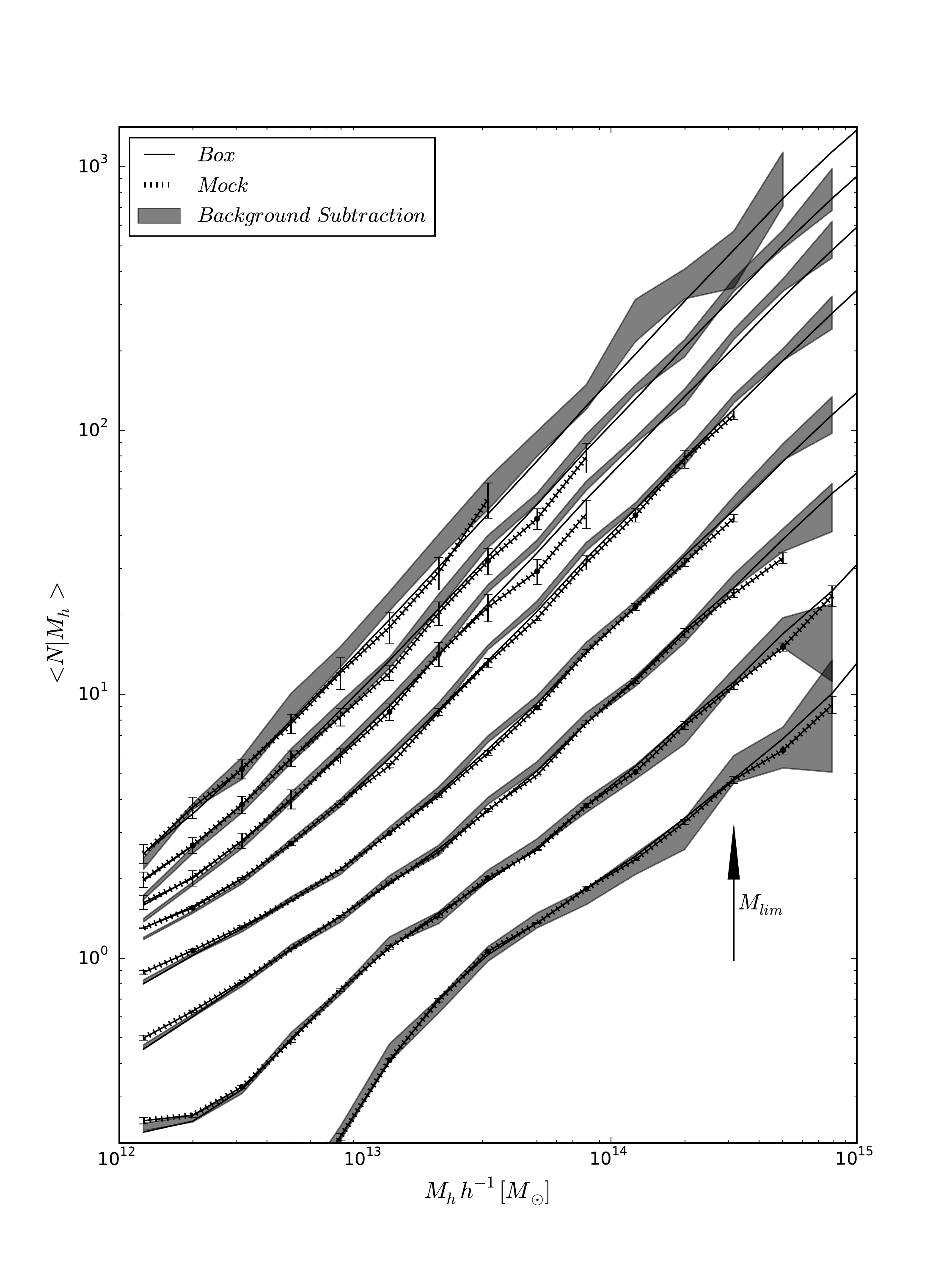}
   \caption{Comparison between HOD obtained by direct counts in volume limited samples 
               (dotted lines), by the background subtraction method (shaded grey bands), 
               and in the simulated box (solid lines).
               From bottom to top, we show the results for $M_{lim} = -21.5$, $-21.0$, $-20.5$, $-20.0$, $-19.0$, $-18.0$, $-17.0$, and $-16.0$.
               The errors are computed using the standard jackknife procedure.
               As can be seen, these results suggest that our method is in good agreement 
               with direct estimation and supports its  robustness.
               It should be taken into account that the proposed method uses both photometric 
               and spectroscopic catalogues, while direct estimation only uses spectroscopic data.}
   \label{fig1}
   \end{figure}

   For each absolute limiting magnitude we apply our method of determining the HOD on 
   the simulated data and compare the values with the HOD obtained by simply counting the number of 
   galaxies in the simulated box. 
   Figure \ref{fig1} shows the results for different absolute limiting magnitudes
   ranging from $M_{lim}=-21.5$ to $M_{lim}=-16.0$. 
   As can be seen, beyond a small overprediction for fainter magnitudes, 
   there is a good agreement between our results (shaded grey bands)
   and those obtained by directly counting galaxies in the simulated box (solid lines) 
   in all ranges of absolute magnitudes.
   We have also included in this figure 
   the HOD corresponding to the volume limited samples to point out the ability of our method
   to achieve higher masses at low luminosities.
   
   As mentioned in section \ref{BS}, it is necessary to restrict the group samples to avoid any bias produced by the fact that 
   massive groups can be observed at higher distances than lower mass groups.
   In the case of the mock catalogue we studied the distribution of the brightest galaxy of each group and we found that 
   for groups with masses $\lesssim 3.16\times 10^{13} M_{\odot}$ it is enough to take a maximum redshift
   of $z=0.11$ given that the tail of the brightest galaxy distribution lies approximately at $M\simeq -19.5$.
   For higher masses a redshift cut is not necessary. Furthermore, for absolute magnitudes fainter than $M=-18$ 
   the group sample should be restricted to a redshift determined by the limiting apparent magnitude of the photometric catalogue 
   (e.g. for the photometric SDSS that have an $r$-band limiting apparent magnitude of $\simeq 21.5$ and for
   $M_{lim}=-17.5$, there is a limiting redshift of $z\simeq 0.15$); even so,
   these samples are much larger than the corresponding volume limited 
   samples (following the previous example, with the corresponding spectroscopic apparent limiting magnitude of $17.77$ 
   and the same limiting absolute magnitude, the resulting maximum redshift is $\simeq 0.04$).
   For this reason, our results cover a wider range of masses than direct 
   measurements; this effect is more pronounced as the absolute magnitudes become fainter,
   as can be seen in figure \ref{fig1}.

\begin{figure}
   \centering
   \includegraphics[width=\hsize]{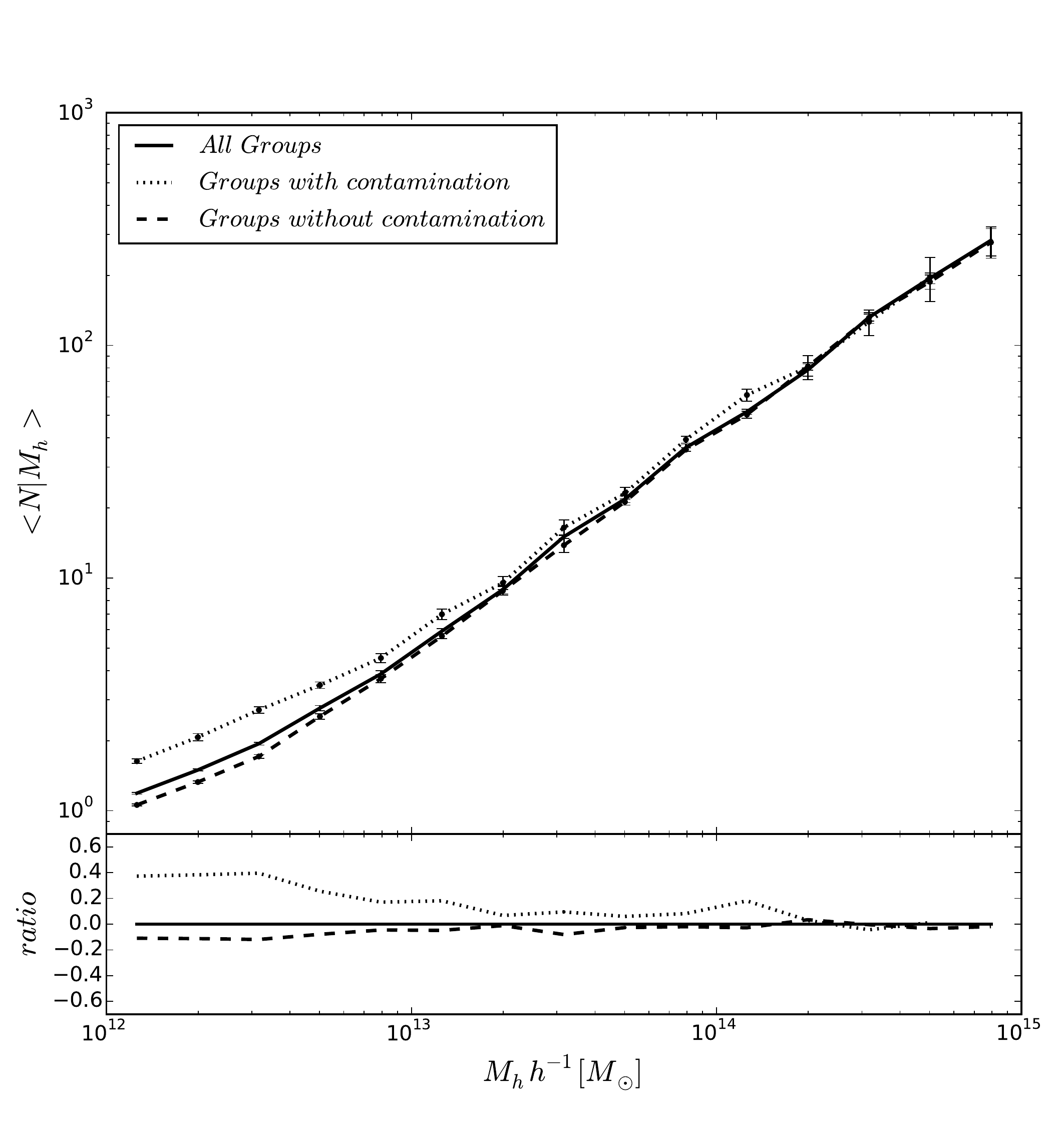}
   \caption{Projection effects in the determination of HOD 
          by the background subtraction method. The dotted line shows the HOD of groups that 
          have, in projections, other groups in the characteristic radius; the dashed line corresponds 
          to the HOD of groups without contamination; and the solid line is the HOD 
          of all groups.
          A standard jackknife procedure is applied to estimate errors.
          For the sake of clarity we only plot the results corresponding to $M_{lim}=-19.0$.
          The bottom panel shows the quotient between the HOD estimated with and without contamination
          and the HOD of all groups.}
   \label{figtest}
  \end{figure}

   A usual critique to the background subtraction method is its   inability to 
   evaluate the contamination produced by background and foreground overdensities, 
   which cannot be distinguished in photometric catalogues. 
   In order to estimate this effect and taking advantage of the mock catalogue, we compute 
   the HOD, distinguishing between groups with and without other groups 
   inside the circle defined by their projected characteristic radius. 
   We find that approximately $17\% $ of groups, by projection effects, have one 
   or more of these contaminating groups. 
   This percentage is homogeneously distributed over the entire range of masses. 
   Figure \ref{figtest} shows the results for all groups (solid line), 
   groups with contamination (dotted line), and without contamination (dashed lines) 
   for $M_{lim}=-19.0$. As can be seen, the HOD of groups with contamination is overestimated 
   for low masses, while groups without contamination reproduce a similar behaviour to 
   the total sample, i.e. contamination effects do not contribute significantly to the 
   background subtraction estimation of total HOD due to the
   small fraction of contaminated groups.
   On the other hand, the HOD does not seem to be affected for higher masses.
   This  is important given that the parameters of the \cite{Yang08} model
   are principally determined by the values of the HOD in this region.  

\begin{figure}
   \centering
   \includegraphics[width=\hsize]{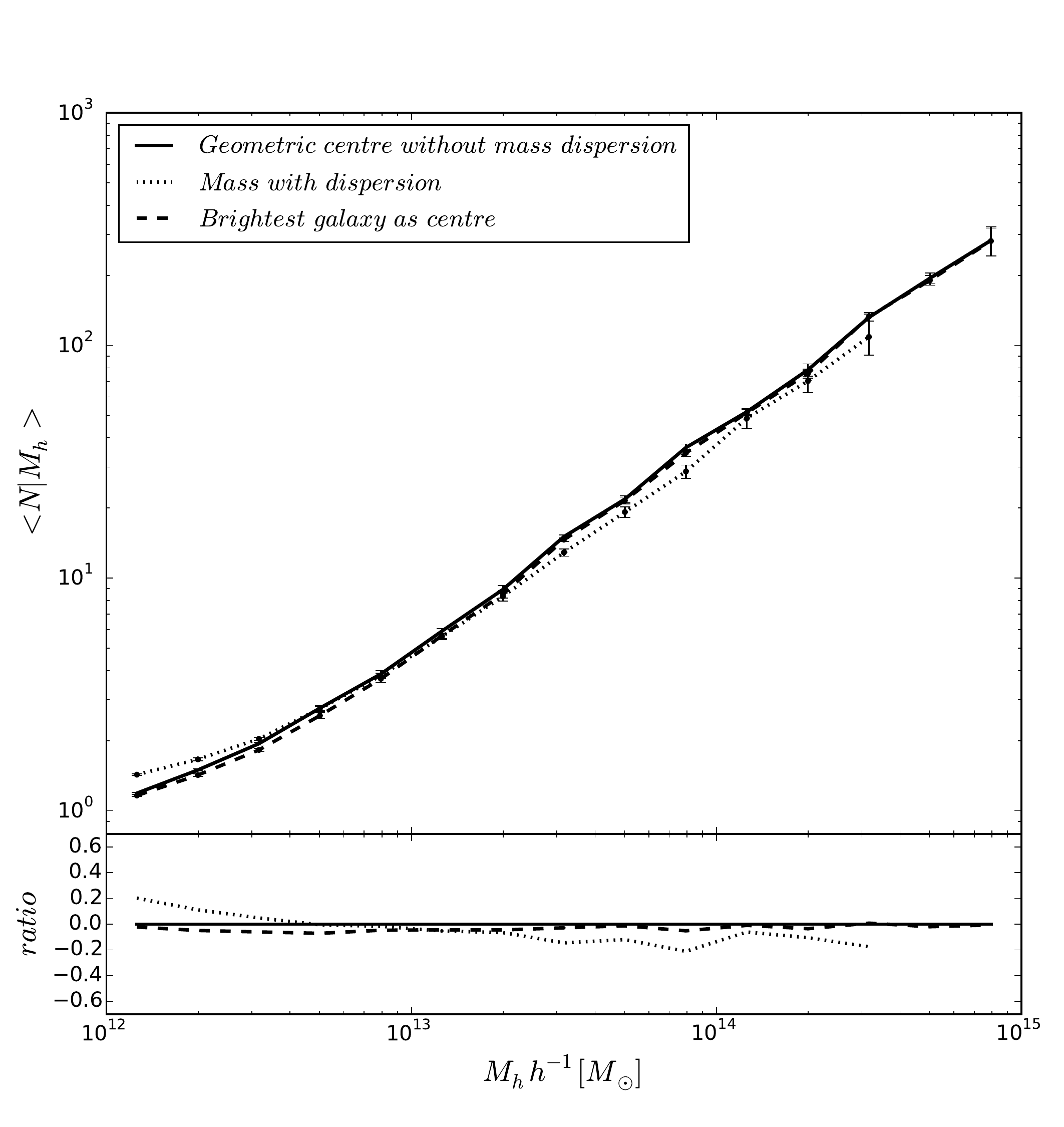}
   \caption{Effects of variations in mass and positions of groups.
            The dotted line corresponds to the groups with an artificial Gaussian dispersion around the simulated halo masses and
            the dashed line shows the HOD of groups using the brightest galaxy as centre.
            The solid line is the HOD of groups in the mock, with geometric centre and masses taken from the simulation, 
            as in previous plots.
            Errors are determined using a standard jackknife procedure.          
            These results correspond to $M_{lim}=-19.0$.
            The bottom panel shows the quotient between the HOD estimations.}
   \label{figtest2}
  \end{figure}

   Another possible source of uncertainty could be the typical error in halo mass estimation.
   To test the impact of these uncertainties on our method, we introduce an artificial Gaussian dispersion
   around the simulated halo masses of 0.2 dex, which is the maximum observed value in the halo mass-number of members relation. 
   This value corresponds to the lower end of mass distribution being much smaller for higher masses. For the sake of clarity, we only show
   in figure \ref{figtest2} the results corresponding to $M_{lim}=-19.0$. As can be seen, the uncertainty in halo masses is not significant in the 
   estimation of HOD.
   
   Throughout this work, we have used the geometric centres of the groups to estimate the HOD. 
   Thus, to complete the evaluation of the method we also consider the impact of a possible offset of the group centres.
   In figure \ref{figtest2} we also include the HOD estimated using the brightest galaxy as centre, 
   demonstrating that the method used to estimate the centres of the groups does not affect our results.  
   
   As a conclusion to this section, we could argue that the HOD estimation using our technique is not affected by projection 
   effects and possible biases in the mass and positions estimates. Furthermore, the results using our method are in excellent 
   agreement with direct count measurements.
   
%______________________________________________________________

   \section{Implementation of the method}
   \label{I}
   
   \subsection{SDSS DR7 HOD}
   
   We now turn to implementing the proposed method to the SDSS DR7 catalogue using the 
   data described in Sect. \ref{2}. Since the catalogue does not cover the whole sky, 
   it is possible that the areas $A_o$ and $A_i$ lie partially out of the catalogue. 
   To take into account the catalogue geometry, we use the routines provided by the SDSSPix software 
   specifically designed for the SDSS geometry (see e.g. \cite{Swanson08}).
   To determine the angular selection function of the survey, we divide the sphere into $\simeq 7,700,000$ equal area pixels. 
   We then compute an initial binary mask by setting to 1 all pixels with at least one galaxy inside, and 0 the remaining pixels.
   Given that the pixel size is much smaller than the mean inter-galaxy separation, we
   smooth the initial mask by averaging each pixel over its $7\times 7$ adjacent neighbours. 
   We compound the angular selection function as the set of all pixels with a value 
   that exceeds a given threshold. To obtain an estimation of $A_i$ and $A_o$ as precisely as possible, we perform
   a second equal area pixelization with a higher resolution (1,962,900,000 pixels).
   The corresponding areas will be proportional to the number 
   of pixels lying in the intersection between the mask and the circles or the
   annuluses, respectively.

   In this case the brightest galaxy distribution is very similar to the mock catalogue, 
   consequently we apply the same restrictions in redshift according to the mass.
   The only difference is for higher masses where the tail of the distribution is at $M\simeq -21.0,$
   which corresponds to a redshift of $z\simeq 0.17$.
   
\begin{figure*}
   \centering
   \includegraphics[width=0.8\hsize]{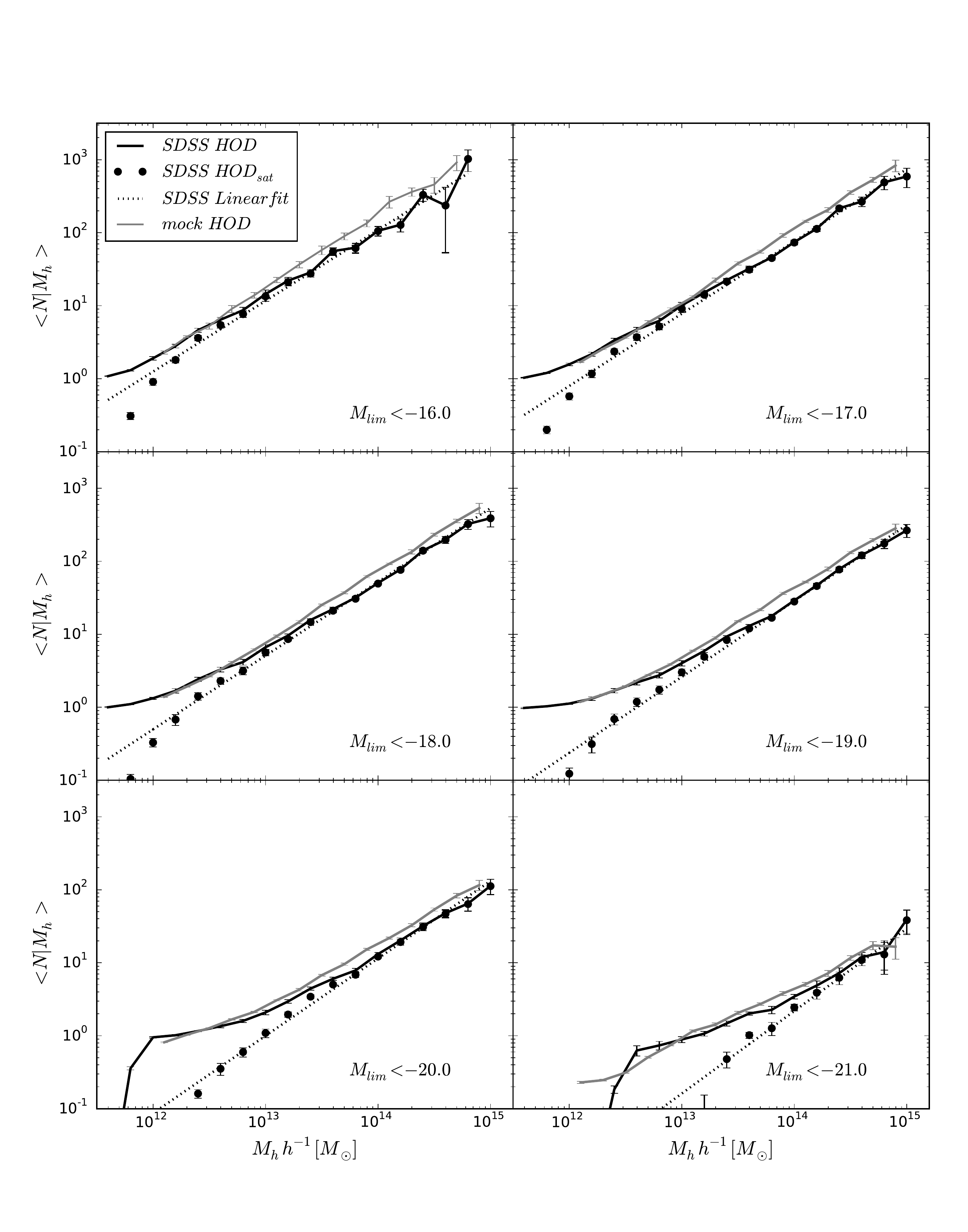}
   \caption{HOD obtained for SDSS DR7 implementing background subtraction method for 
            different limiting absolute magnitude ($M_{lim} = -16.0, -17.0, -18.0,-19.0,-20.0$ and 
            $-21.0$ as indicated in each panel). 
            We show in solid lines the total HOD while solid circles show the satellite 
            distribution. Dotted lines correspond to the best linear fit of satellite 
            distribution.
            Just for comparison, we included the occupation corresponding to the synthetic catalogue estimated in the previous section (gray lines).
            Error bars are computed using the standard jackknife procedure.}
   \label{fig2}
  \end{figure*}

   In figure \ref{fig2} we show the   HOD obtained for six ranges of limiting absolute magnitudes
   as indicated in each panel. The expected behaviour of these statistics can be seen, i.e.
   the decrease in the number of galaxies  at brighter magnitudes and 
   the characteristic decay of the HOD described by \cite{Zheng05} (only visible in the bottom panels).
      
  Although the results of the semi-analytical model cannot be compared directly with those for the Sloan
  because the \cite{Guo11} semi-analytical model was not specifically made to reproduce this statistic,
  a comparison between SDSS DR7 HOD (solid lines) and  mock HOD (grey lines) in figure \ref{fig2} 
  shows the semi-analytic results to be slightly higher than the corresponding values from the SDSS. 
  This discrepancy suggests that the HOD could be used to improve the semi-analytical model.
    
   \subsection{Comparison with previous results}
   
\begin{figure}
    \centering
    \includegraphics[width=\hsize]{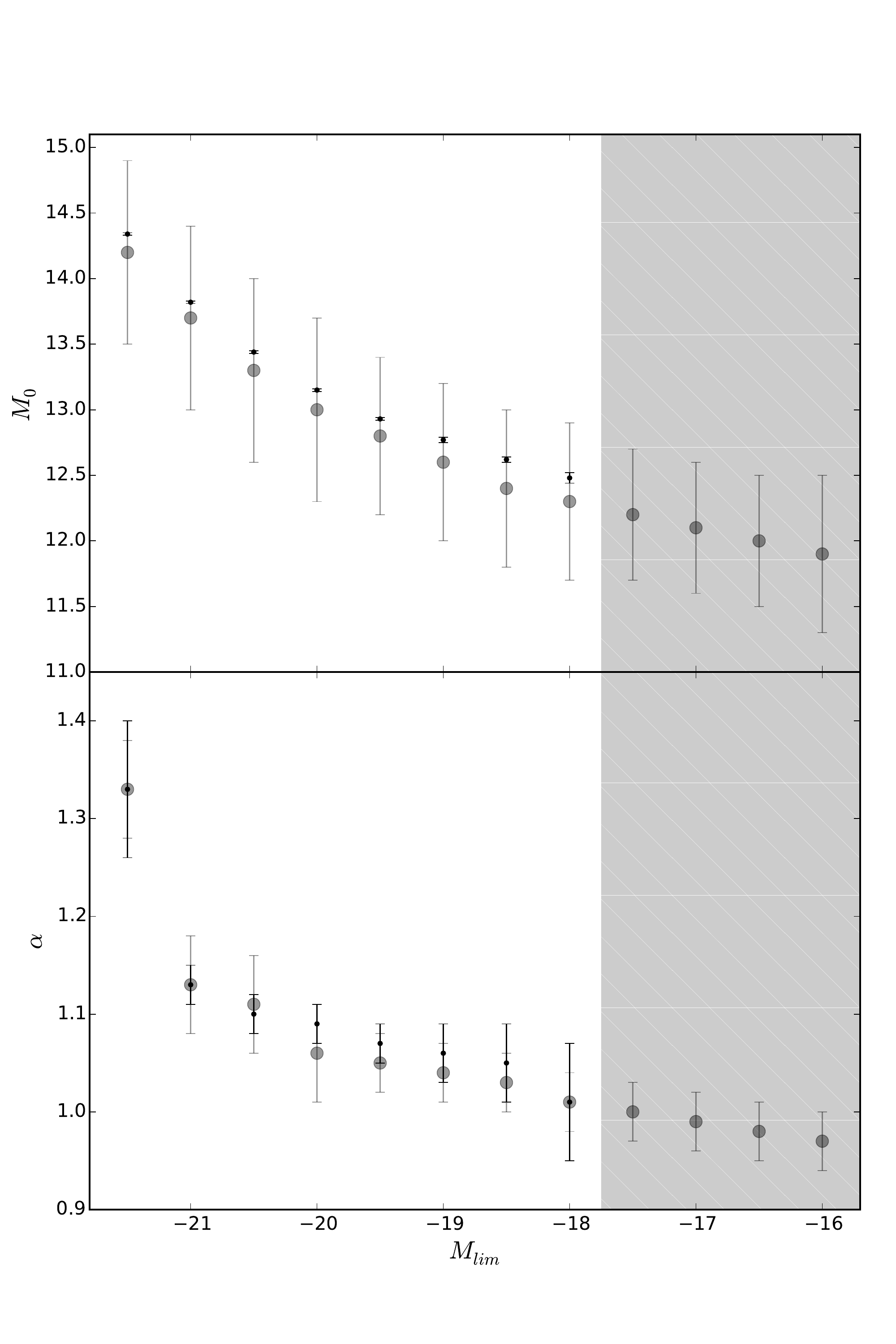}
    \caption{Parameters of HOD for satellite galaxies (see equation \ref{rel}).
             Top and bottom panels show $M_0$ and $\alpha$ parameters, respectively, 
             as a function of the limiting magnitude. Our results are shown in grey 
             circles while black circles correspond to results of \citet{Yang08}.
             There is a good agreement between both results, furthermore our method 
             allows to extend the range of absolute magnitude to fainter values up to 
             $M_{lim}=-16.0$ without losing statistical strength (shaded area). 
             Parameters and error bars correspond to the best values obtained following the $\chi ^2$ Minimization Method.}.
   \label{fig3}
   \end{figure}

   To compare our results with those of \cite{Yang08} for SDSS data release 4, we use the parameterization of 
   the satellite galaxies distribution proposed by these authors:
   \begin{equation}
   \label{rel}
   \langle N_s|M_h \rangle = \left(\frac{M_h}{M_0}\right)^\alpha
   \end{equation}
   were $M_0$ is the characteristic mass of a halo above which there is on average
   at least one satellite, and $\alpha $ is the power-law index.
   In figure \ref{fig2} we have included the fit for each range of limiting absolute magnitude (dotted lines) and
   in figure \ref{fig3} we plot the best fit parameters as a function of the limiting absolute magnitude (grey circles) together 
   with the corresponding to \cite{Yang08}. As can be seen, there is an excellent agreement between our results and previous results. 
   Furthermore, this figure makes clear an important feature of our method: the ability to estimate the HOD for fainter absolute 
   magnitudes than the previous works, the shaded area in the figure shows the new values obtained with this technique. 
   For completeness and to clarify, the information presented in figure \ref{fig3} is also compiled in table~1.

 \begin{table}
  \centering  
  \caption{Comparison between parameters of HOD for satellite galaxies obtained by \citet{Yang08} and by our method.}
  \begin{tabular}{@{}ccccc@{}}
  
  \hline
&\multicolumn{2}{c}{ \cite{Yang08}} 
&\multicolumn{2}{c}{Background subtraction}\\
  $M_{lim}$ &  $M_0$ & $\alpha$  &  $M_0$ & $\alpha$\\ 
 \hline
$-16.0$ & $------$       & $------$     &$11.9\pm 0.6$&$0.97\pm 0.03$\\
$-16.5$ & $------$       & $------$     &$12.0\pm 0.5$&$0.98\pm 0.03$\\
$-17.0$ & $------$       & $------$     &$12.1\pm 0.5$&$0.99\pm 0.03$\\
$-17.5$ & $------$       & $------$     &$12.2\pm 0.5$&$1.00\pm 0.03$\\
$-18.0$ & $12.48\pm 0.04$&$1.01\pm 0.06$&$12.3\pm 0.6$&$1.01\pm 0.03$\\
$-18.5$ & $12.62\pm 0.02$&$1.05\pm 0.04$&$12.4\pm 0.6$&$1.03\pm 0.03$\\
$-19.0$ & $12.77\pm 0.02$&$1.06\pm 0.03$&$12.6\pm 0.6$&$1.04\pm 0.03$\\
$-19.5$ & $12.93\pm 0.01$&$1.07\pm 0.02$&$12.8\pm 0.6$&$1.05\pm 0.03$\\
$-20.0$ & $13.15\pm 0.01$&$1.09\pm 0.02$&$13.0\pm 0.7$&$1.06\pm 0.05$\\
$-20.5$ & $13.44\pm 0.01$&$1.10\pm 0.02$&$13.3\pm 0.7$&$1.11\pm 0.05$\\
$-21.0$ & $13.82\pm 0.01$&$1.13\pm 0.02$&$13.7\pm 0.7$&$1.13\pm 0.05$\\
$-21.5$ & $14.34\pm 0.01$&$1.33\pm 0.07$&$14.2\pm 0.7$&$1.33\pm 0.05$\\
\hline
\end{tabular}
\end{table}

\begin{figure}
    \centering
    \includegraphics[width=\hsize]{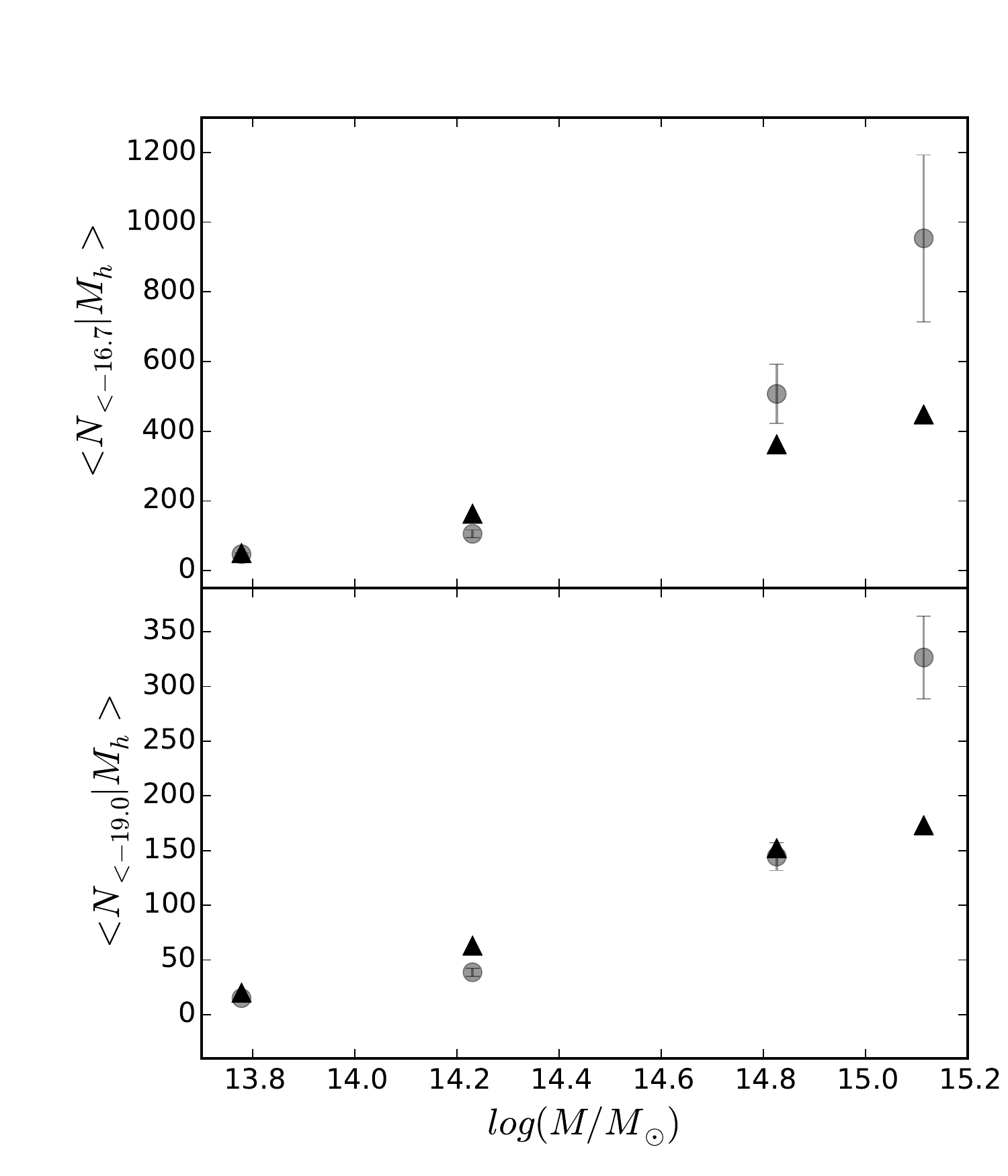}
    \caption{Comparison between galaxy abundances of four observed clusters taken from \cite{Weinmann} (triangles) and the corresponding results obtained with discount method (circles) in mass bins centred in these clusters.
             Here, we use masses of $6 \times 10^{13} M_{\odot}$ for Fornax, $1.7\times 10^{14} M_{\odot}$ for Virgo, $6.7 \times 10^{14} M_{\odot}$ for Perseus and $1.3 \times 10^{15} M_{\odot}$ for Coma.
             Top panel show the abundances for $M_{lim} = -16.7$ and bottom panel for $M_{lim} = -19.0$.}
   \label{fig5}
   \end{figure}

\cite{Weinmann} collects information for four well studied galaxy clusters (Fornax, Virgo, Perseus, Coma) that can be confronted directly with our results.
Figure \ref{fig5} shows the abundance of these clusters for two ranges of limiting absolute magnitudes taken from \cite{Weinmann} and the HOD in mass bins centred in the 
mass of these clusters obtained applying our method to the \cite{Yang12} group sample and the photometric version of the SDSS catalogue. The size of the errors 
associated with the largest mass is a logical outcome of that the function was estimated using 9 groups in the case of $M_{lim} = -16.7$ and 14 groups in 
the case of $M_{lim} = -19.0$.
The abundances of Fornax, Virgo and Perseus seem to be in agreement with the mean number of galaxies for systems of the same masses, meanwhile the Coma cluster 
shows a population significantly smaller.  While this seems to be in contraposition with the results of \cite{Weinmann}, it should be noted that these authors make 
the comparison with a single cluster taken from the semi-analytical model of  \cite{Guo11}, which does not necessarily represent the behaviour of the corresponding 
mass media. 

%______________________________________________________________
\section{Conclusions}
\label{C}

   In this paper we present a statistical approach to estimating the HOD based on the 
   background subtraction technique that combines a spectroscopic galaxy group catalogue with
   a photometric galaxy survey. 
   Given the nature of the photometric catalogue our method is able to deal with large volumes, 
   consequently the HOD can be estimated in a wider range of masses even when faint absolute magnitudes are considered. 
   A further consequence of the use of a photometric galaxy catalogue is the ability
   to  measure   weaker magnitudes than was possible when  using spectroscopic 
   information alone.

   Our method was tested using mock catalogues constructed from a semi-analytic model of 
   galaxy formation \citep{Guo10} applied on top of the Millennium I Simulation.
   We show that the background subtraction technique is able to recover the
   HOD resulting from the model and from the mock catalogue using a direct count 
   in volume limited samples as described by \cite{Yang08}.
   Additionally, we studied projection effects as a source of possible bias
   as well as the impact of errors on the mass and centre estimates.
   We find that only $17\%$ of groups have one or more overlapping projection groups 
   resulting in a statistically unnoticeable distortion with respect to the model; 
   in addition, the introduction of variations in the mass and positions of the groups does not significantly affect our results.

   We obtained the HOD of SDSS DR7 in twelve ranges of limiting absolute magnitudes
   which expand from $-16.0$ to $-21.5$ in Sloan r-band. 
   This estimation was performed using the groups catalogue
   provided by \cite{Yang12} and the SDSS DR7 photometric catalogue.
   Our results are consistent with those obtained by \cite{Yang08}, and
   given the nature of our method, the samples of groups used to estimate the HOD are 
   deeper and more numerous, allowing us to extend the range of masses beyond the limits 
   explored in previous works and also to achieve absolute magnitudes as faint as $M_{lim}-16.0$.
   
   The proposed technique is able to combine two
   catalogues, one of groups and the other  of galaxies, that occupy the same sky area
   independently of the origin of the group catalogue and the photometric 
   bandpass of the galaxy survey. Its features make it particularly useful for studying 
   the behaviour of the HOD as a function of different properties of groups or galaxies
   and will allow us to fully exploit the benefits of the new generations of galaxy catalogues.
   Possible applications could be the study of the population of dwarf galaxies (in the range of $M_{lim}=-16.0$ to $-19.0$)
   or constraints on semi-analytical models.
   
   Finally, we will use the results of this work to extend the anisotropic halo model developed by \cite{Sgro}
   in order to evaluate the ability of this model to reproduce observational results.

\begin{acknowledgements}

   We would like to thank to Dante Paz for reading the manuscript and helpful suggestions. 
   
   This work was partially supported by the Consejo Nacional de Investigaciones Científicas y Técnicas (CONICET), the Secretaría de
   Ciencia y Tecnología (SeCyT) and the Observatorio Astronómico de Córdoba (OAC) both from the Universidad Nacional de Córdoba.  
\end{acknowledgements}

%-------------------------------------------------------------------

\end{document}